\documentclass[
      draft            % use draft while you are working on the paper
  ,numberedheadings % uncomment this option for numbered sections
  ]
  {aipproc}

\layoutstyle{8x11single}

\begin{document}

\title{Origin of Anomalous Xe-H in Nanodiamond Stardust}


\author{K.-L. Kratz}{
  address={Max-Planck-Institut f\"ur Chemie, Otto-Hahn-Institut, D-55128 Mainz, Germany}
  ,altaddress={Fachbereich Chemie, Pharmazie und Geowissenschaften, Universit\"at Mainz, Mainz, Germany\\
   $^\dagger$ Department of Physics, Univ. of Notre Dame, Notre Dame, IN 4656556, USA}
%  ,altaddress={Department of Physics, Univ. of Notre Dame, Notre Dame, IN 4656556, USA} 
% additional visiting address
}

\author{K. Farouqi}{
  address={Max-Planck-Institut f\"ur Chemie, Otto-Hahn-Institut, D-55128 Mainz, Germany}
  ,altaddress={Zentrum f\"ur Astronomie der Universit\"at Heidelberg, D-69120 Heidelberg, Germany}
}

\author{O. Hallmann}{
  address={Max-Planck-Institut f\"ur Chemie, Otto-Hahn-Institut, D-55128 Mainz, Germany}
%  ,altaddress={<author1 address>} % additional visiting address
}

\author{B. Pfeiffer}{
  address={II. Physikalisches Institut, Univ. Gie\ss en, D-35392 Gie\ss en, Germany}
%  ,altaddress={<author1 address>} % additional visiting address
}

\author{U. Ott}{
  address={Max-Planck-Institut f\"ur Chemie, Otto-Hahn-Institut, D-55128 Mainz, Germany}
  ,altaddress={Univ. of West Hungary, H-9700 Szombathely, Hungary} % additional visiting address
} 

\begin{abstract}
 Still today, the nucleosynthesis origin of Xe-H in presolar nanodiamonds is far from understood.
Historically, possible explanations were proposed by a secondary ``neutron-burst'' process
occurring in the He- or C/O-shells of a type-II supernova (SN-II), which are, however, not
fully convincing in terms of modern nucleosynthesis conditions. Therefore, we have investigated 
Xe isotopic abundance features that may be diagnostic for different versions of a classical,
primary r-process in high-entropy-wind (HEW) ejecta of core-collapse SN-II. We report here
on parameter tests for non-standard r-process variants, by varying electron abundances
(Y$_e$), ranges of entropies (S) and expansion velocities (V$_{exp}$) with their correlated
neutron-freezeout times ($\tau$(freeze)) and temperatures (T$_9$(freeze)). From this study,
we conclude that a best fit to the measured Xe-H abundance ratios $^i$Xe/$^{136}$Xe
can be obtained with the high-S ``main'' component of a ``cold'' r-process variant.
\end{abstract}

\maketitle

\section{Introduction}

Apart from the historical ``bulk'' Solar System (S.S.) isotopic abundances (N$_\odot$)
\cite{B2FH,cameron,coryell,lodders2009} and
the elemental abundances measured for metal-poor halo stars (for a review, see e.g. 
\cite{sneden2008,spite}),
meteoritic grains of stardust, which survived from times before the S.S. formed (see, e.g.
\cite{clayton2004,lodders2005,zinner})
represent the third group of ``observables'' crucial for our understanding of the various
nucleosynthesis processes in stars. Within the realm of meteoritc grains, attempts to explain
the origin of nanodiamonds have not progressed as much as the understanding of other types
of stardust, such as for example SiC and oxide grains. A major problem is the small size
(average $\simeq$2.6 nm), which does not permit classical single-grain isotopic measurements.
Therefore, ``bulk'' samples  (i.e. millions of tiny nanodiamond grains) have to be analyzed.
In nanodiamonds, peculiar isotopic features have been observed in several trace elements,
which seem to suggest a connection to SNe. These include Xe-HL \cite{huss1994,ott2010} 
(with enhancements
of light (L) p- and heavy (H) r-isotopes relative to N$_\odot$), Kr-H 
and 
Pt-H \cite{ott2010,ott-m} (where the heavy (H) isotopes are enhanced), and Te-H 
\cite{richter1998} 
(with a clear overabundance of the
``r-only'' isotopes). So far, nucleosynthesis processes suggested to account for the H
r-process enhancements include ``neutron-burst'' scenarios of secondary nature 
\cite{clayton1989,meyer2000,rauscher},
as well as a regular primary r-process augmented by an ``early'' separation between final,
stable isotopes and radioactive isobars formed during the $\beta$-decay back to stability
from the initial precursors in the assumed r-process path \cite{ott1996}. Since both 
earlier scenarios
are not fully convincing, as a possible alternative we have initiated a concerted effort to look
for isotopic features that may be diagnostic for non-standard variants of an r-process in
high-entropy-wind (HEW) ejecta of core-collapse (cc) SNe (for details of the HEW scenario,
see e.g. \cite{woosley1994,takahashi1994,freiburghaus,farouqi2009}).

\section{THE HEW R-PROCESS MODEL}

The basic nucleosynthesis mechanisms for elements beyond Fe by slow (s-process) and
rapid (r-process)  captures of neutrons have been known for a long time 
\cite{B2FH,cameron,coryell}.
However, the search for a robust r-process production site has proven difficult. Still
today, all proposed scenarios not only face problems with astrophysical conditions, but
also with the necessary nuclear-physics input for very neutron-rich isotopes. Among
the various suggested sites, the neutrino-driven or high-entropy wind (HEW) of cc-SNe
is one of the best studied mechanisms for a full primary r-process already in the early
Universe (see, e.g. \cite{woosley1994,takahashi1994,freiburghaus,farouqi2009}).
Nevertheless, also for this attractive scenario even in the
most sophisticated hydrodynamical models the neutrino-driven HEW has
been found to be proton-rich (electron fraction Y$_e$ $\ge$ 0.5) during its entire life,
thus precluding a rapid neutron-capture process  (see, e.g. \cite{fischer,hude}). 
However, recent
work on charged-current neutrino interaction rates (see, e.g. \cite{roberts}), seems to revive 
the HEW scenario by predicting that Y$_e$ changes from an initially proton-rich value
to moderately neutron-rich conditions with minimal values of Y$_e$ $\simeq$ 0.42.  \\
Therefore, in the light of these recent findings, to us it appears to be justified of further
using our parameterized, dynamical HEW approach, based on the initial model of
Freiburghaus et al. \cite{freiburghaus}, which assumes adiabatically expanding 
homogenious mass
zones with different entropy (S) yields. This code has been steadily improved until today,
for example to implement  a better mathematical treatment for the tracking of the
$\beta$-decaying nuclei back to stability with time-intervals small enough to consider
late recaptures of previously emitted $\beta$-delayed neutrons, up to the longer-lived
precursors 55-s $^{87}$Br and 24-s $^{137}$I. \\
As has been outlined in \cite{farouqi2009}, in our HEW model the overall wind ejecta 
represent model-inherently weighted superpositions of S (and probably also Y$_e$) 
components, where the
main astrophysical parameters are correlated via the ``r-process strength formula''
(Y$_n$/Y$_{seed}$ $\simeq$ k$_{SN}$ $\times$ V$_{exp}$ $\times$ (S/Y$_e$)$^3$). Following our
traditional approach to find a possible explanation of the N$_{r,\odot}$ ``residuals''
\cite{lodders2009,bisterzo},
we have started the present study using our standard parameter combination of
Y$_e$ = 0.45, V$_{exp}$ = 7500 km/s and the full S-range of 20 $\le$ S $\le$ 280 k$_B$/baryon.
As indicated in Table~\ref{tab1}, under these conditions we have a ``hybrid'' r-process type with a
neutron-freezeout temperature of T$_9$(freeze)  $\simeq$ 0.82, which is reached at a time
$\tau$(freeze) $\simeq$ 138 ms after the r-process seed formation. This r-process variant
reaches the maximum abundance at the A $\simeq$ 130 peak at S $\simeq$ 195 for
Y$_n$/Y$_{seed}$ $\simeq$ 35 with the top of the peak at A = 128. Thereafter, we have
systematically investigated the whole astrophysics parameter space as functions of electron
fraction (0.40 $\le$ Y$_e$ $\le$ 0.495), expansion velocity (1000 $\le$ V$_{exp}$ $\le$ 30,000
km/s) and entropy ranges. In Table~\ref{tab1}, we show some relevant parameters for the 
formation
of the A $\simeq$ 130 peak with Y$_e$ = 0.45 and the corresponding full S-ranges for
different possible r-process variants, ranging from a ``hot, fast'' version with an r-process
boulevard at moderate distance from the stability line, up to a ``cold, rapid'' process with
r-progenitors far from stability. \\

\begin{table}
\begin{tabular}{llllll}
\hline
V$_{exp}$ & S  [k$_B$/baryon]  &  maximum    &     $\tau_r$ [ms]   &  T$_9$  &        r-process \\

 [km/s] &   \multicolumn{2}{c}{2$^{nd}$ r-peak} & \multicolumn{2}{c}{at neutron-freezeout} & variant \\
\hline
   1000       &         370       &        A=130      &       560        &       1.23     &     hot \& fast \\  

   2000       &         298        &                   &           335       &        1.10    &  \\

   4000       &         235       &                    &           200       &        0.98     &  \\                                 

   6000        &        210        &       A=128        &     160       &        0.86        &     hybrid \\

   8000        &        190        &                    &         130      &          0.82    &      (standard) \\ 

10,000        &        175         &                 &           110       &         0.79     &   \\

15,000        &        155          &                &             93      &          0.66     &  \\

20,000        &        140       &         A=126     &        78       &         0.60      &    cold \& rapid  \\  

\hline
\end{tabular}
\caption{Relevant r-process parameters for the formation of the A $\simeq$ 130 abundance
peak with Y$_e$ = 0.45, the respective full S-ranges and Y$_n$/Y$_{seed}$ = 35 for
different r-process variants. All HEW calculations were performed with the theoretical
nuclear-physics input based on the mass model ETFSI-Q \cite{ETFSI-Q,moller2003} and the local 
QRPA \cite{arndt} improvements mentioned in the text.
}
\label{tab1}
\end{table}

\section{REPRODUCTION OF THE Xe-H ABUNDANCES}

\begin{figure}
  \includegraphics[height=.25\textheight,angle=0]{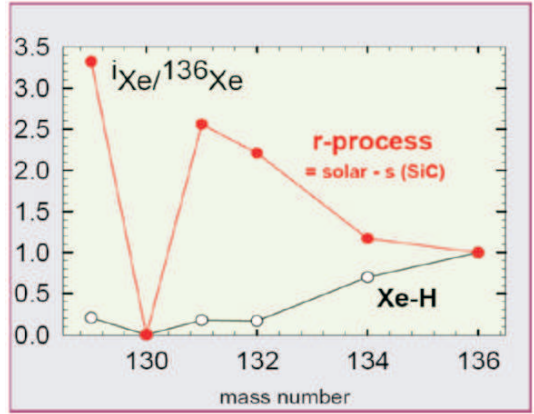}
\caption{
Isotopic composition of Xe-H compared to S.S. r-process Xe, adopted
from  
\cite{clayton1989}. The abundance ratios are normalized to ``r-only''
$^{136}$Xe.}
\label{fig1}
\end{figure}

In the nanodiamond stardust samples, the Xe-H abundance pattern is given as
ratios relative to the assumed ``r-only'' isotope $^{136}$Xe, and have (somewhat
model dependent) values of $^{129,131,132,134}$Xe/$^{136}$Xe = 0.207,
0.178, 0.167, 0.699 
\cite{ott2010,ott-m}.
As is shown in Fig.~\ref{fig1}, these values are clearly different from the N$_{r,\odot}$
abundance ratios of 3.37, 2.55, 2.24, 1.18  
\cite{bisterzo}. The low
measured values of the three lighter (r+s) Xe isotopes indicate a significant 
reduction of their abundances by factors of 16.3, 14.3 and 13.4, respectively,
whereas the abundance ratio of the two heavy ``r-only'' isotopes (96.2 \% r
$^{134}$Xe and 99.95 \% $^{136}$Xe) is only moderately changed by a
factor 1.7. In the historical ``neutron-burst'' model of 
\cite{clayton1989,meyer2000},
this Xe abundance pattern has been obtained by a neutron-
capture shift of an s-processed seed composition in an explosive shell-burning
scenario. In the HEW scenario, however, such a strong abundance shift to a
peak at N = 82 $^{136}$Xe, with its main r-progenitor N = 86 $^{136}$Sn, 
is very unlikely. It would require a drastic change of the position and height of
the N$_{r,\odot}$ peak with its classical, most abundant N = 82 ``waiting-point''
isotopes $^{128}$Pd, $^{129}$Ag and $^{130}$Cd (see, e.g. \cite{arndt}), 
$\beta$-decaying to the
stable isobars $^{128}$Te, $^{129}$Xe and $^{130}$Te. Hence, if a primary
rapid neutron-capture process would be the nucleosynthesis origin of the peculiar
Xe-H pattern, possibilities within more likely non-standard r-process variants
have to be investigated.

\begin{figure}
  \includegraphics[height=.25\textheight,angle=0]{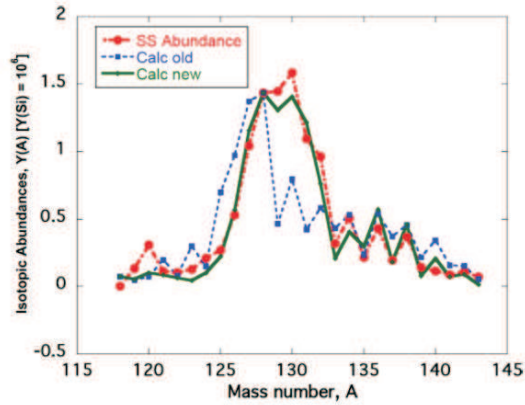}
 \caption{
Observed S.S.-r abundance ``residuals'' (N$_{r,\odot}$ =
N$_\odot$ - N$_{s,\odot}$) 
\cite{lodders2009,bisterzo}
(filled circles connected with a
dot-dash line) along with the isotopic abundances calculated as described in the
text, using older values for $\beta$-decay properties (filled squares connected
with a dashed line) and recently updated $\beta$-decay half-lives and delayed
neutron-branching ratios (filled diamonds connected with a solid line). The
calculated values are normalized to the S.S.-r abundance of $^{128}$Te.
}
\label{fig2}
\end{figure}

However, before studying the individual, and later also possible combined effects
of the main HEW parameters Y$_e$, V$_{exp}$ and S, we first have to verify that
our r-process model is able to reproduce the isotopic pattern of the whole
N$_{r,\odot}$ peak region. In this context, our experimental information on
direct r-process progenitor isotopes as well as a detailed understanding of the
nuclear-structure development in the N = 82 magic-shell region (see, e.g. 
\cite{arndt}) is essential. Only with both, realistic astrophysical and
nuclear-physics parameters, a satisfactory agreement between S.S.-r observations
and our HEW calculations can serve as a ``basis'' for the later interpretation of
abundance patterns deviating from the standard N$_{r,\odot}$ pattern. In Fig.~\ref{fig2},
we show two fits of the 120 $\le$ A $\le$ 140 peak region, both with the nuclear-physics 
input based on the ETFSI-Q mass model, but the one with older theoretical
$\beta$-decay properties, and the other with updated half-lives and P$_{xn}$
values, as described in Arndt et al. \cite{arndt}.
With the latter, new
nuclear-physics input, for the above mass range the mean abundance ratio of
N$_{r,calc}$/N$_{r,\odot}$ $\simeq$ 0.90. To our knowledge, this is the best
agreement one has been able to achieve so far for the region of the 2$^{nd}$ r-peak.

\begin{figure}
 \includegraphics[height=.25\textheight,angle=0]{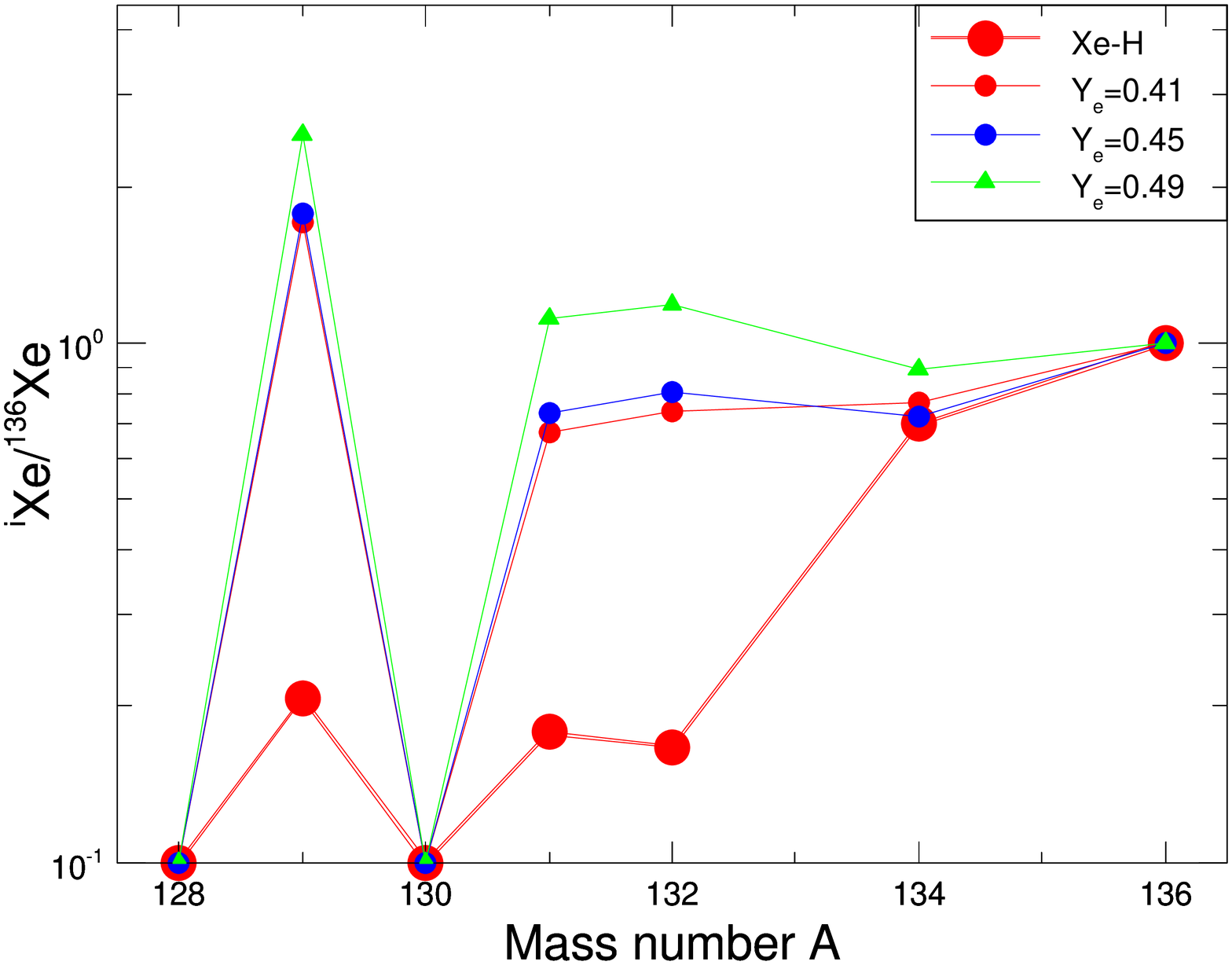}
\caption{
HEW predictions of $^i$Xe/$^{136}$Xe abundance ratios for r-process ejecta with 
different electron fractions Y$_e$, with a constant expansion velocity V$_{exp}$ = 
7500 [km/s] and the respective full entropy ranges. The case Y$_e$ = 0.49 corresponds 
to only slightly neutron-rich conditions as observed for metal-deficient ``r-poor'' 
halo stars; Y$_e$ = 0.45 represents moderately neutron-rich ejecta as observed for 
``r-rich'' stars; and Y$_e$ = 0.41 corresponds to strongly neutron-rich conditions. 
The Xe-H data are included for comparison. For further details, see text.}
\label{fig3}
\end{figure}

We now want to check the possible influence of the neutron-richness of the 
r-process ejecta on the $^i$Xe/$^{136}$Xe abundance ratios. Therefore, 
we vary the HEW parameter Y$_e$, i.e. the electron fraction, in the range 
0.41 $\le$ Y$_e$ $\le$ 0.49, while keeping the wind expansion velocity 
V$_{exp}$ = 7500 [km/s] constant and taking the respective full entropy 
ranges (see also Table~\ref{tab1}). In Fig.~\ref{fig3}, we show three 
typical results in comparison with the Xe-H measurements 
\cite{ott2010}. Here, the case Y$_e$ = 0.49 corresponds 
to only slightly neutron-rich conditions, as observed for r-process poor 
``Honda-type'' halo stars (e.g. HD 122563 
\cite{honda}); Y$_e$ = 0.45 represents 
moderately neutron-rich conditions for r-process enriched, metal-deficient 
``Sneden-type'' stars (e.g. CS 22892-052 
\cite{sneden2003}); and Y$_e$ = 0.41 corresponds 
to strongly neuron-rich, ``Cayrel-type'' stars (e.g. CS 31082-001 
\cite{hill}), which 
show a ``main'' r-process component with a so-called ``actinide boost''. As 
can be seen from Fig.~\ref{fig3}, the measured low Xe abundance ratios are not 
met by varying the electron fraction parameter alone. But, the trend shows 
that high Y$_e$ values may be less likely than the lower ones. Therefore, 
we conclude that the Xe-H pattern seems to favor neutron-rich HEW ejecta.

\begin{figure}
 \includegraphics[height=.25\textheight,angle=0]{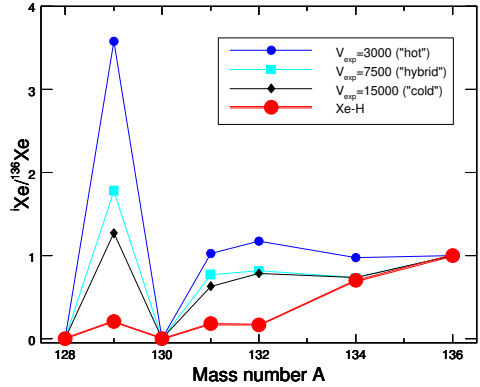}
\caption{
HEW predictions for $^{i}$Xe/$^{136}$Xe abundance ratios for
neutron-rich r-process ejecta with different expansion velocities V$_{exp}$,
with a constant electron fraction Y$_e$ = 0.45 and the respective full entropy
ranges.
The case V$_{exp}$ = 7500 [km/s] represents our
``standard'' parameter combination for a ``hybrid'' r-process; 
V$_{exp}$ =
3000 [km/s] corresponds to a fast, ``hot'' r-process; and V$_{exp}$ = 15,000
[km/s] represents a rapid, ``cold'' r-process variant.
For comparison, again the Xe-H data 
are included. For further details see text and \cite{farouqi2009}.
}
\label{fig4}
\end{figure}

In a second step, we vary the individual parameter of the HEW expansion 
velocity V$_{exp}$ in the range 1000 $\le$ V$_{exp}$ $\le$ 30,000 [km/s], 
while now keeping the electron fraction Y$_e$ = 0.45 constant and taking 
again the full respective entropy ranges (see also Table~\ref{tab1}). 
In Fig.~\ref{fig4}, we show 
again three typical results in comparison with the Xe-H data 
\cite{ott2010}. Here, the case 
V$_{exp}$ = 7500 [km/s] corresponds to our ``standard'' HEW parameter 
combination for a ``hybrid'' r-process 
\cite{farouqi2009}; V$_{exp}$ = 3000 [km/s] represents 
a moderately fast, ``hot'' r-process; and V$_{exp}$ = 15,000 [km/s] corresponds 
to a rapid, ``cold'' r-process variant (for further details, see 
\cite{farouqi2009}). As is evident 
from this figure, again the measured Xe-H abundance ratios are not met by 
varying the ejection velocities of the r-process ejecta alone. However, the trend 
in this case indicates that relatively high V$_{exp}$ values may be more likely 
than low ones. What we do not show here, is our further result that very high 
V$_{exp}$ $\ge$ 20,000 [km/s] may, however, be excluded too. Therefore, we conclude 
from this HEW parameter choice that the Xe-H pattern seems to favor rapid, 
neutron-rich, ``cold'' r-process  ejecta.

The last HEW parameter to ``play'' with in our present study is the range of the 
superimposed entropy (S) components. From 
\cite{farouqi2009} we know that with increasing 
S, we can distinguish between three different primary, rapid nucleosynthesis 
components. In the lowest S-range, where the HEW has no or not yet enough 
``free neutrons'' (Y$_n$/Y$_{r-seed}$ $\le$ 1) we have a charged-particle 
process. For somewhat higher S-values, a relatively low fraction of ``free neutrons'' 
(1 $\le$ Y$_n$/Y$_{r-seed}$ $\le$ 20 -- 30) occurs, which leads to a ``weak'' neutron-capture 
process that produces r-matter up to the rising wing of the A $\simeq$ 130 
peak. And finally, for high S-values (and moderately low Y$_e$; see above) 
high neutron fractions of up to Y$_n$/Y$_{r-seed}$ $\simeq$ 150) can be obtained, 
under which conditions a ``main'' r-process component becomes possible. This 
r-component forms a robust S.S.-r like REE ``pygmy peak'', the full A $\simeq$ 195 
peak, and may reach up to the Th and U r-chronometer isotopes. \\ 
 %FIGURE 5. 
\begin{figure}
  \includegraphics[height=.25\textheight,angle=0]{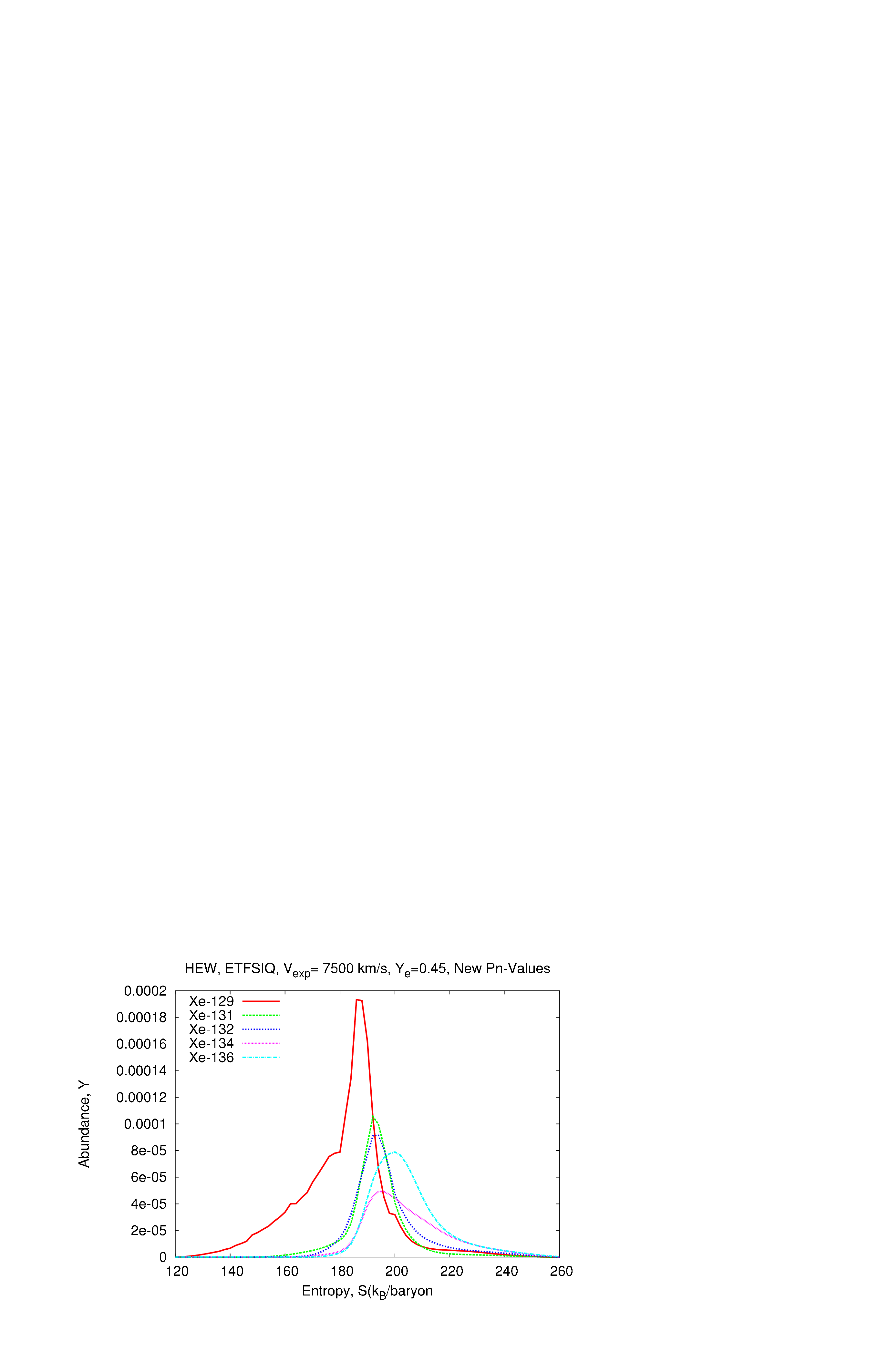}
\caption{
Xe abundances (Y) of HEW ejecta for our ``standard'' parameter
combination of Y$_e$ = 0.45 and V$_{exp}$ = 7500 [km/s] as a function of
entropy S [k$_B$/baryon]. The Xe isotopes are 
produced in slightly different S-ranges, where $^{129}$Xe abundance maximum occurs 
at a lower entropy that those for the heavier isotopes.
For futher details, see text.
}
\label{fig5}
\end{figure}
% $\dots$ \\
Given this development of S and Y$_n$/Y$_{r-seed}$, it is immediately evident that 
the r-process Xe isotopes between $^{129}$Xe and $^{136}$Xe are predominantly 
produced by the ``main'' component, which (quite logically) suggests to cut out the 
low-S ranges of the charged-particle and the ``weak-r'' components. Furthermore, 
a closer look to the production of the individual Xe isotopes as a function of S shows, 
that they are formed under slightly different conditions. The most significant differences 
occur for the lightest r-isotope $^{129}$Xe, which is placed in the left part of the top 
of the 2$^{nd}$ r-peak, and the heaviest r-nuclide $^{136}$Xe, which sits already beyond 
the right, decreasing wing of the peak, close to the beginning of the REE region. For our 
``standard'' HEW parameter combination of Y$_e$ = 0.45 and V$_{exp}$ = 7500 [km/s], 
this situation is shown in Fig.~\ref{fig5}. From this figure, it can be seen that $^{129}$Xe is 
mainly produced at entropies of S $\le$ 190 (by the initial, classical N = 82 r-process 
``waiting-point'' nuclide $^{129}$Ag), whereas the heavier Xe isotopes need higher 
entropies.

\begin{figure}
 \includegraphics[height=.25\textheight,angle=0]{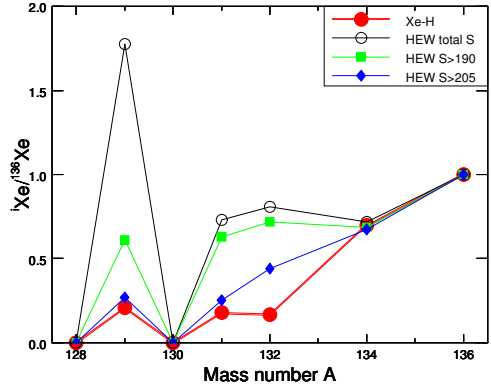}
\caption{
HEW predictions of $^{i}$Xe/$^{136}$Xe abundance ratios  for
different entropy ranges, with constant ``standard'' r-process conditions of
Y$_e$ = 0.45 and V$_{exp}$ = 7500 [km/s], compared to the measured Xe-H data. 
The results obtained with the cuts of the low-S ranges, in particular the 
``best fit'' for S $\ge$ 205, represent the signature of a moderately neutron-rich 
``hybrid, main'' r-process variant. 
For comparison, again the Xe-H data are included. For further discussion, see text.
}
\label{fig6}
\end{figure}

This picture now suggests to check the effect of selecting only the high-S components 
out of the full S-range. Again, for our ``standard'' HEW parameter combination for Y$_e$ 
and V$_{exp}$, this is shown in Fig.~\ref{fig6}. We see that in particular the 
high ``overabundance''
of $^{129}$Xe can be nicely reduced by the requested order of magnitude, while the 
abundance ratio of the two ``r-only'' isotopes $^{134}$Xe and $^{136}$Xe remains 
practically unaffected. From this result, our near-final conclusion is that the measured 
Xe-H pattern seems to be the signature of a moderately neutron-rich, ``hybrid, main'' 
(or even ``strong'') r-process variant.

\begin{figure}
  \includegraphics[height=.25\textheight,angle=0]{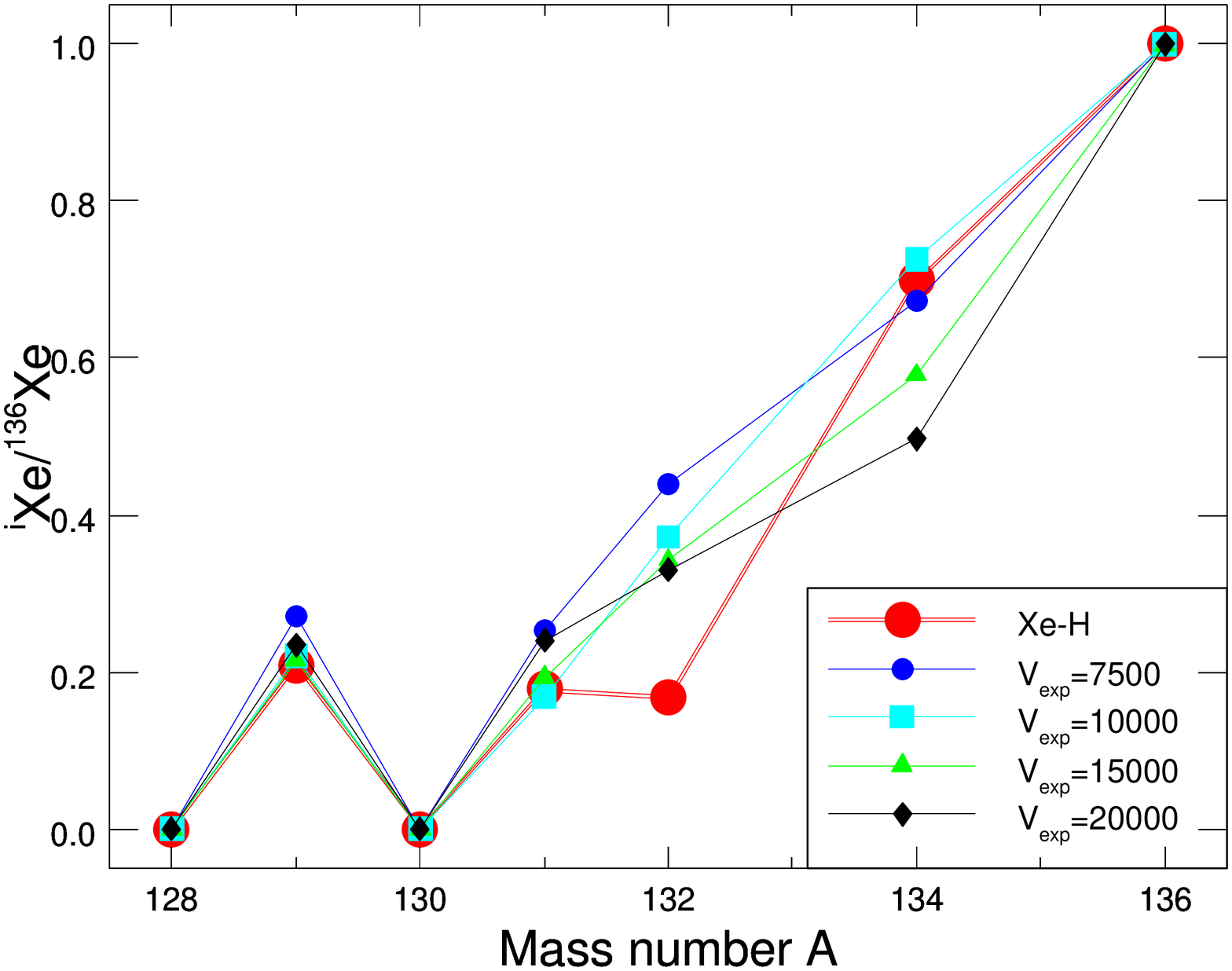}
\caption{
HEW predictions of $^i$Xe/$^{136}$Xe abundance ratios,
combining the ``best'' individual astrophysics parameters for Y$_e$ $\simeq$ 0.43 
and the respective S-cuts with different choices of V$_{exp}$. 
%for a ``cold,rapid'' r-process variant. 
 ``Best fit'' conditions are obtained for somewhat
higher expansion velocities than our standard value, in the range 10,000 $\le$ 
V$_{exp}$ $\le$ 15,000 [km/s]. For further
discussion, see text.
}
\label{fig7}
\end{figure}

After we have so far studied the effects of the main individual HEW parameters, and 
have determined their possible optimum values, at the very end we can combine these 
values and compare the respective abundance predictions to the measured Xe-H 
pattern. This is shown in our final Fig.~\ref{fig7} for a limited number of ``best fit'' 
combinations of Y$_e$ 
$\simeq$ 0.43, the respective low-S cuts and a range of reasonable, but not too
high  V$_{exp}$ values.  
Similar to our result already obtained from the parameters used in Fig.~\ref{fig6}, we conclude 
that the Xe-H abundance ratios indicate its presolar formation by a ``rapid, cold, main'' 
r-process variant. 
%(see also Fig.~\ref{fig6}). 
The measured abundance ratios of 
$^{129,131,134}$Xe/$^{136}$Xe are well reproduced within 20 to 30 \%. Only the 
value of $^{132}$Xe/$^{136}$Xe remains too high by about a factor 2.5.  It is interesting 
to note in this context, that also the historical ``neutron-burst`` model 
%[12,13]
\cite{clayton1989,meyer2000} predicted a 
somewhat too high abundance of $^{132}$Xe.  \\

Finally, it is worth to be mentioned that preliminary results for platinum  indicate that 
the HEW conditions found to be favorable for Xe-H can also account for the reported Pt-H 
in presolar diamonds \cite{ott2010,ott-m}. With Xe-H lying in the A $\simeq$ 130 N$_{r,\odot}$ 
peak and 
Pt-H in the A $\simeq$ 195 peak, these cosmochemical samples with their isotopic abundance 
patterns can provide constraints on the astrophysical conditions for the production of a 
full, ``main'' r-process, which cannot be deduced with this sensitivity by the elemental 
abundance patterns of metal-poor halo stars.

%\begin{theacknowledgments}

%\end{theacknowledgments}

\bibliographystyle{aipproc}   % if natbib is available

%\endinput

\end{document}